\documentclass[aps,amsmath,pre]{revtex4}
\usepackage{graphicx}
\usepackage{float}
\begin{document}

\title{Degree Correlations in a Dynamically Generated Model Food Web}
\author{
Per Arne Rikvold
}
\affiliation{
Department of Physics and Center for Materials Research and Technology\\
Florida State University, Tallahassee, FL 32306-4350, USA
}

\begin{abstract}
We explore aspects of the community structures generated by a
simple predator-prey model of biological coevolution, using 
large-scale kinetic Monte Carlo simulations. The model accounts 
for interspecies and intraspecies competition for resources, as well
as adaptive foraging behavior. It produces a metastable low-diversity
phase and a stable high-diversity phase. The structures and joint 
indegree-outdegree distributions of the food webs generated in the 
latter phase are discussed.
\end{abstract}

\maketitle              

\section{Introduction and Model}
\label{sec:int}

Biological evolution and ecology 
involve nonlinear interactions between large numbers of units and 
have recently become popular topics among statistical and computational
physicists \cite{DROS01}. However, many models used by
physicists are unrealistic to the extent of attracting 
little attention from biologists. Here, we introduce 
a somewhat more realistic model of the dynamics of a predator-prey system 
and explore some aspects of the resulting food-web structures. 

Recently, we developed simplified models of biological macroevolution
\cite{RIKV03,RIKV06}, in which
the reproduction rates in an individual-based population 
dynamics with nonoverlapping generations provide the mechanism for 
selection between several interacting species. New species 
enter the community through point
mutations in a haploid, binary ``genome" of length $L$.
The potential species are identified by the index $I \in [0,2^L-1]$. 
(Typically, only $\mathcal{N}(t) \ll 2^L$
species are present in the community at any time $t$.) At the end of each 
generation, each individual of species $I$ gives birth to a fixed number 
$F$ of offspring with probability $P_I$ before dying, or dies
without offspring with probability $(1-P_I)$. Each offspring may
mutate into a different species with a small probability $\mu$. Mutation 
consists in flipping a randomly chosen bit in the genome. 

Here, we consider a model with modified population
dynamics that include competition between different
predators that prey on the same species, as well as a satiation
effect for predators with abundant prey. 
Consistent with our previous work \cite{RIKV06}, 
the central quantity of the model is 
an {\it antisymmetric interaction matrix\/} $\bf M$ 
representing predator-prey interactions. Thus, $M_{IJ} > 0$ and $M_{JI} < 0$ 
means that $I$ is a predator and $J$ its prey, and vice versa. 
The elements of the upper triangle of $\bf M$ are drawn randomly from a 
symmetric distribution over $[-1,+1]$ and kept constant during the whole 
simulation (quenched randomness). 
A constant, $R>0$, represents an external resource. The 
ability of species $I$ to utilize $R$ is $\eta_I$, which 
with probability $c_{\rm prod} \ll 1$ is chosen to be a random number 
uniform on $(0,+1]$ (i.e., $c_{\rm prod}$ is the 
proportion of potential producer species). 
Species with $\eta_I =0$ are consumers. 
The population size of species $I$ is $n_I$. 

Interspecies competition is modeled by defining the number of
individuals of species $J$ that are available as prey for $I$, corrected for
competition from other predator species, as 
\begin{equation}
\hat{n}_{IJ} = \frac{n_I M_{IJ}}{\sum_L^{{\rm pred}(J)} n_L M_{LJ}} n_J \;,
\label{eq:neff}
\end{equation}
where $\sum_L^{{\rm pred}(J)}$ runs over all $L$ 
such that $M_{LJ} > 0$. Thus, 
$\sum_I^{{\rm pred}(J)} \hat{n}_{IJ} = n_J$.
Analogously, we define the competition-adjusted
external resources available to a producer species $I$ as 
$\hat{R}_I = R {n_I \eta_I}/{\sum_L n_L \eta_L}$.
With these definitions, the total,
competition-adjusted resources available to species $I$ are 
\begin{equation}
\hat{S}_I = \eta_I \hat{R}_I + \sum_J^{{\rm prey}(I)} M_{IJ} \hat{n}_{IJ} 
\;,
\label{eq:SI}
\end{equation}
where $\sum_J^{{\rm prey}(I)}$ runs over all $J$ such that $M_{IJ} > 0$. 

The {\it functional response\/} of species $I$ with respect to $J$, 
$\Phi_{IJ}$, is the rate at which an individual of species $I$ consumes
individuals of $J$ \cite{DROS01B,KREB01}. 
For ecosystems consisting of a single pair of
predator and prey, or a simple chain from a bottom-level
producer through intermediate species to a top predator, the most
common forms of functional response are due to Holling \cite{KREB01}.
For more complicated food webs, several
functional forms have been proposed recently, 
\cite{DROS01B,SKAL01,DROS04,MART06} 
but there is as yet no agreement about a standard form.
Here, we model intraspecies competition by a ratio-dependent \cite{RESI95} 
Holling Type II \cite{KREB01} form due to Getz \cite{GETZ84},
\begin{equation}
\Phi_{IJ} = \frac{M_{IJ} \hat{n}_{IJ}}{\lambda \hat{S}_I + n_I} \;,
\label{eq:PhiIJ}
\end{equation}
where $\lambda \in (0,1]$ is the metabolic efficiency of converting
prey biomass to predator offspring. 
Analogously, the functional response of a producer species 
$I$ toward the external resource $R$ is 
$\Phi_{IR} = {\eta_I \hat{R}_{I}}/[{\lambda \hat{S}_I + n_I}]$.
The total consumption rate for an individual of species $I$ is therefore 
\begin{equation}
C_I = \Phi_{IR} + \sum_J^{{\rm prey}(I)} \Phi_{IJ} 
= \frac{\hat{S}_I}{\lambda
\hat{S}_I + n_I}
=
\left\{
\begin{array}{lll}
\hat{S}_I/n_I & \mbox{for} & \lambda \hat{S}_I \ll n_I \nonumber\\
1/\lambda     & \mbox{for} & \lambda \hat{S}_I \gg n_I 
\end{array}
\right.
\;.
\label{eq:CI}
\end{equation}
The birth probability 
is assumed to be proportional to the consumption rate, 
$B_I = \lambda C_I \in [0,+1]$,
while the probability that an individual of $I$ 
avoids death by predation until attempting to reproduce is 
\begin{equation}
A_I = 1 - \sum_J^{{\rm pred}(I)} \Phi_{JI} \frac{n_J}{n_I} \;.
\label{eq:AI}
\end{equation}
The reproduction probability for an individual of species $I$
is $P_I(t) = A_I(t) B_I(t)$. 

\begin{figure}[htb]
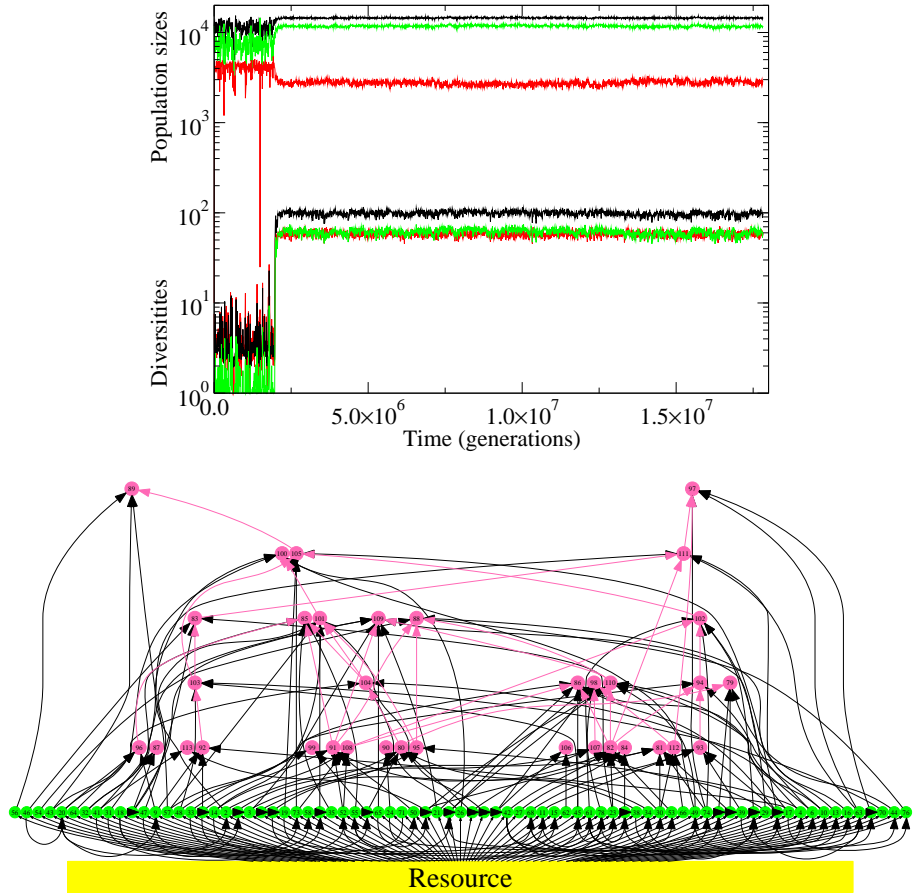

\begin{center}
\vspace*{0.6truecm}
\includegraphics[angle=0,width=.55\textwidth]{timsADFig.eps}\\
\vspace{0.2truecm}
\includegraphics[angle=0,width=0.8\textwidth]{graph8.eps}
\end{center}
\caption[]{
{\bf (Top)}
Time series of 
populations (upper curves) and diversities (lower curves) for the 
adaptive model. All species (black), producers (light gray, green online), and 
consumers (dark gray, red online). 
{\bf (Bottom)}
A representative food web. 
The producers are shown just above the external resource. 
Above them are the consumers, connected to producers by black arrows and 
to other consumers by gray (pink online) arrows. 
Arrows point from prey to predator.
}
\label{fig:adap}
\end{figure}
As the model is defined above, species forage
indiscriminately over all available resources, with the output only
limited by competition. Also, there is an implication that an
individual's total foraging effort increases proportionally with
the number of species to which it is connected by a positive  
$M_{IJ}$. A more realistic picture would be that an individual's
total foraging effort is constant and can either be divided
equally, or concentrated on richer resources. This is known as
adaptive foraging. While one can 
go to great length devising optimal foraging
strategies \cite{DROS01B,DROS04}, we here
only use a simple scheme, in which individuals of $I$ show a
preference for prey species $J$, based on the interactions and
population sizes (uncorrected for interspecies competition) and given by 
\begin{equation}
g_{IJ} = \frac{M_{IJ}n_J}{\eta_I R + \sum_K^{{\rm prey}(I)} M_{IK} n_K}
\;,
\label{eq:gij}
\end{equation}
and analogously for $R$ by  
$g_{IR} = {\eta_{I} R}/[{\eta_I R + \sum_K^{{\rm prey}(I)} M_{IK} n_K}]$.
The total foraging effort is thus 
$g_{IR} + \sum_J^{{\rm prey}(I)} g_{IJ} = 1$. 
These preference factors are used to modify the reproduction 
probabilities by replacing all occurrences of $M_{IJ}$
by $M_{IJ} g_{IJ}$ and of $\eta_I$ by $\eta_I g_{IR}$ in 
Eqs.~(\ref{eq:neff}) -- (\ref{eq:PhiIJ}). 

\section{Numerical Results}
\label{sec:Sim1}

\begin{figure}[htb]
\begin{center}
\vspace*{0.2truecm}
\includegraphics[angle=0,width=.50\textwidth]{Hist_Correl.eps}
\hspace{0.5truecm}
\includegraphics[angle=0,width=.44\textwidth]{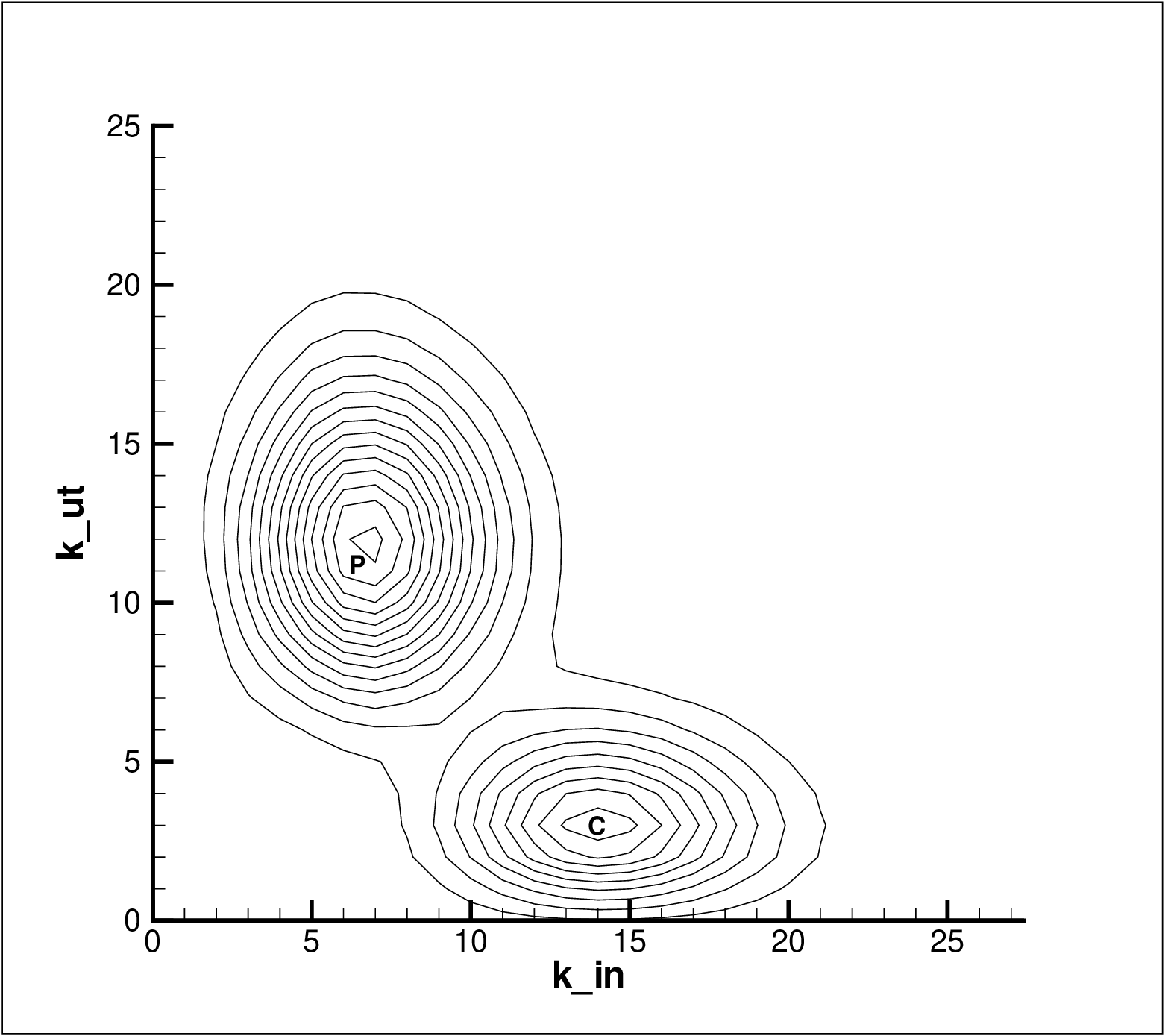}
\end{center}
\caption[]{
{\bf (Left)}
Correlations between indegree (number of prey) and outdegree (number of 
predators) for full and core communities and 17 empirical communities 
\protect\cite{RIKV06}. 
{\bf (Right)}
Joint probability distribution for indegree and outdegree in core communities. 
The simulation results in both parts were averaged over 327\,680 communities. 
}
\label{fig:adap2}
\end{figure}
We simulated the model over
$2^{24} = 16\,777\,216$ generations (plus $2^{20}$
generations ``warm-up") for the
following parameters: $L=21$ 
($2^{21} = 2\,097\,152$ potential
species), $R=16\,000$, $F=2$,  
$\mu = 10^{-3}$, $c_{\rm prod} =0.05$, interaction matrix $\bf M$ with
connectance $C = 0.1$ and nonzero elements with a symmetric, 
triangular distribution over $[-1,+1]$, and $\lambda = 1.0$. 
We ran five independent runs, each starting from a population of 
100 randomly chosen producer species. 

Time series of diversities (effective numbers of species)
and population sizes for one run are shown in Fig.~\ref{fig:adap}(top). 
To filter out noise from low-population,  
unsuccessful mutants, the diversity is defined as the exponential
Shannon-Wiener index \cite{KREB89}. This is the exponential function 
of the information-theoretical entropy of the population distributions,
$\rho_I(t) = n_I(t) / N_{\rm tot}(t)$ for the case of all species,
and analogously for the producers and consumers separately. 

Without adaptive foraging, the system flips randomly 
between a phase with a diversity near ten, and a 
phase of one or a few producer species with a very low population of
many unstable consumer species \cite{RIKV09}. 
Adaptive foraging produces a striking change in the 
dynamics. There is now a metastable low-diversity phase, which 
gives way at a random time to a stable
high-diversity phase with much smaller fluctuations. 
As seen in Fig.~\ref{fig:adap}(top), the switch-over is quite abrupt.

A representative community food web is shown 
in Fig.~\ref{fig:adap}(bottom). This is a ``core community," 
extracted from the full community by retaining only species with $n_I >1$ 
that also existed 256 generations earlier. 
Here, every consumer species preys on at least 
one producer species, thus there are only two trophic levels. 
(Only links with $|M_{IJ}| \ge 0.5$ are included.)

Histograms of the correlation coefficient between a species' numbers of 
prey (indegree) and predators (outdegree) are shown in 
Fig.~\ref{fig:adap2}(left). The correlations are strongly negative in both 
the simulated full and core communities and also in the majority of the 
17 empirical communities considered in Ref.~\cite{RIKV06}. These negative 
correlations are explained by the joint indegree-outdegree distribution 
shown in Fig.~\ref{fig:adap2}(right): 
producers (P) have low average indegree and high average outdegree, 
while consumers (C) show the opposite behavior.

\section{Conclusions}
\label{sec:conc}

We have introduced a model for the biological coevolution 
of predators and prey, based on the ecological concept of functional response.
When adaptive foraging is included in the model, it has a dynamically 
stable phase of relatively high diversity. The indegree and 
outdegree of a species are negatively correlated, which is 
explained by the observation that producers have low indegree and high 
outdegree, while consumers have high indegree and low outdegree. 
Our model demonstrates the high degree of complexity that can be 
produced, even by simple models of biological evolution and ecology. 

\section*{Acknowledgments}

Supported in part by NSF Grant Nos.\ DMR-0444051 and DMR-0802288.




\end{document}